\documentclass[12pt]{iopart} 

\usepackage{graphicx,subcaption}
\usepackage{comment}
\usepackage{hyperref}
\usepackage{amssymb,bm,calc}
\begin{document}

\title {Required sensitivity to search the neutrinoless double beta decay in $^{124}Sn$}

\author{Manoj Kumar Singh,$^{1, 2} \mbox{\large$^{\ast}$}$ Lakhwinder Singh,$^{1, 2}$ Vivek Sharma,$^{1, 2}$ Manoj Kumar Singh,$^{1}$ Abhishek Kumar,$^{1}$ Akash Pandey,$^{1}$ Venktesh Singh,$^{1}\mbox{\large$^{\ast}$}$ Henry Tsz-King Wong$^{2}$}

\address{$^{1}$ Department of Physics, Institute of Science, Banaras Hindu University, Varanasi 221005, India.}
\address{$^{2}$ Institute of Physics, Academia Sinica, Taipei 11529, Taiwan.}
\ead{$^{\ast}$ manojsingh.physics@gmail.com}
\ead{$^{\ast}$ venkaz@yahoo.com}
\date{\today}

\begin{abstract}
  \textbf{T}he \textbf{IN}dia’s \textbf{TIN} (TIN.TIN) detector is under development in the search for 
  neutrinoless double-$\beta$ decay (0$\nu\beta\beta$) using 90\% enriched $^{124}$Sn 
  isotope as the target mass. This detector will be housed in the upcoming underground facility of the \textbf{I}ndia 
  based \textbf{N}eutrino \textbf{O}bservatory. We present the most important 
  experimental parameters that would be used in the study of required sensitivity 
  for the TIN.TIN experiment to probe the neutrino mass hierarchy. The sensitivity of the TIN.TIN 
  detector in the presence of sole two neutrino double-$\beta$ decay (2$\nu\beta\beta$) 
  decay background is studied at various energy resolutions. The most optimistic and 
  pessimistic scenario to probe the neutrino mass hierarchy at 3$\sigma$ sensitivity level and 
  90\% C.L. is also discussed.
  
\vspace{0.6in}  
\end{abstract}


\maketitle

\section{Introduction}
\label{intro}
Neutrinoless double-$\beta$ decay (0$\nu\beta\beta$) is an interesting 
venue to look for the most important question whether neutrinos have Majorana or 
Dirac nature. During the last two decades, the discovery of non-zero neutrino mass 
and mixing with various sources gives new motivation for more sensitive searches of 0$\nu\beta\beta$. 
In fact, the observation of 0$\nu\beta\beta$ would not only establish the 
Majorana nature of neutrinos, but also provide a measurement of effective mass and probe 
the neutrino mass hierarchy. Furthermore, this is the only proposed process which has potential to allow 
the sensitivity of the absolute mass scale of neutrino below 100 meV. There is no exact 
gauge symmetry associated with lepton number, therefore there is no fundamental reason 
why lepton number should be conserved at all levels~\cite{ADGO, GSTE}. The lepton number 
violates by two units in the case of 0$\nu\beta\beta$. This distinctive feature 
together with CP (charge parity) violation supports the exciting possibility that neutrino 
plays an important role in the matter-antimatter asymmetry in the early universe. \par

The experimental search for 0$\nu\beta\beta$ is an attractive field of nuclear and particle 
physics. There are several isotopes available which energetically allow 0$\nu\beta\beta$ 
process, only 35 of them are stable and have their experimental importance~\cite{RHEN}.
Several experiments are focusing on different isotopes via utilizing various detector 
techniques such as GERDA (GERmanium Detector Array)~\cite{AGOS}, MAJORANA (Majorana Demonstrator)~\cite{AALS} and 
CDEX (China Dark matter EXperiment) with $^{76}$Ge enrich high purity Ge detectors~\cite{WANG}; EXO 
(Enriched Xenon Observatory)~\cite{ALBE} and KamLandZen (Kamioka Liquid Scintillator Antineutrino Detector) 
with liquid $^{136}$Xe time projection chambers~\cite{GAND}; and CUORE (Cryogenic Underground Observatory 
for Rare Events) with $^{130}$Te bolometric detectors~\cite{ALDU}. 
The next-generation experiments with tonne scale detectors such as LEGEND ($^{76}$Ge) (MAJORANA + 
GERDA)~\cite{ABGR, SCHW}, nEXO ($^{136}$Xe)~\cite{JBAL}, NEXT ($^{136}$Xe)~\cite{MART}, 
CUPID ($^{130}$Te)~\cite{GWAN}, SuperNEMO ($^{82}$Se, $^{150}$Nd)~\cite{RBPA}, 
AMoRE ($^{100}$Mo)~\cite{VALE}, COBRA ($^{116}$Cd)~\cite{JEBE}, CANDLES-III ($^{48}$Ca)~\cite{TIID}, 
SNO$^{+}$ ($^{130}$Te)~\cite{VLOZ}, TIN.TIN ($^{124}$Sn)~\cite{VNAN}, MOON ($^{100}$Mo)~\cite{TSHI} 
and LUMINEU ($^{100}$Mo)~\cite{EARM} have been proposed. Some of them will 
start data taking over the next few years and others are under construction phase. These large number 
of experiments, reveals the enthusiasm of the scientists working in this field world-wide. \par
\begin{figure} 
  \centering
  \includegraphics[width=8.5cm]{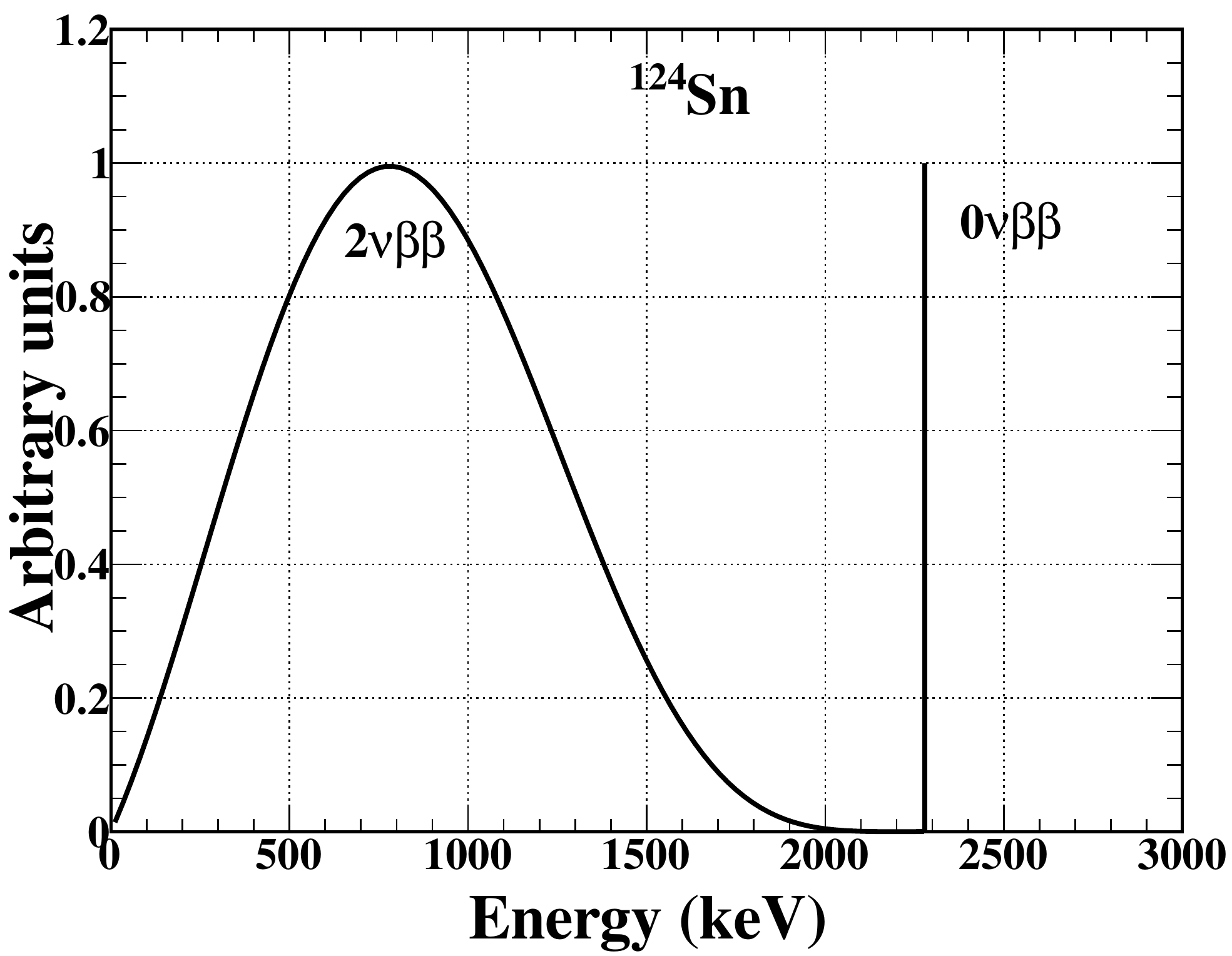}
  \caption{Summed energy spectrum of two electrons emitted in 2$\nu\beta\beta$ 
and 0$\nu\beta\beta$ decay modes of $^{124}$Sn.}
  \label{fig:Spectrum}
\end{figure}

The two neutrino double-$\beta$ decay (2$\nu\beta\beta$) is a second order weak process, 
in which two neutrons simultaneously transfer into two protons by emitting two electrons 
and two anti-neutrinos within the same nucleus~\cite{OCRE}
\begin{equation}
^N_Z A_{\beta\beta} ~ \rightarrow ~ _{Z+2}^{N-2}A ~ + ~ 2 e^- ~ + ~ 2\bar{\nu}_{e}.  ~~
\label{eq:2nubb}
\end{equation} 
The energy spectrum of 2$\nu\beta\beta$ process has a continuous spectrum, ending at 
a well-defined end point which is determined by the Q$_{\beta\beta}$-value of the process, 
as depicted in Fig. \hyperref[fig:Spectrum]{1}. The 2$\nu\beta\beta$ decay follows the 
conservation of lepton number and also allowed by the standard model~\cite{OCRE}. 
In the case of 0$\nu\beta\beta$ process, no neutrino is emitted and both electrons 
carry the full energy equal to the Q$_{\beta\beta}$-value of the transition. 
Indeed, being the energy of the recoiling nucleus negligible due to its high mass. 
Therefore, the experimental signature of 0$\nu\beta\beta$ is 
characterized by a monoenergetic peak at the Q$_{\beta\beta}$-value which relies just 
on the detection of the two emitted electrons. 
\begin{equation}
^N_Z A_{\beta\beta} ~ \rightarrow ~ _{Z+2}^{N-2}A ~ + ~ 2 e^{-}  ~~
\label{eq:0nubb}
\end{equation}

\par

The TIN.TIN (\textbf{T}he \textbf{IN}dia’s \textbf{TIN}) detector is under development 
in search of 0$\nu\beta\beta$ in $^{124}$Sn isotope. The TIN.TIN 
detector will use the cryogenic bolometer technique in closely packed module 
structure arrays~\cite{VNAL}. This experiment will be housed at the \textbf{I}ndia 
based \textbf{N}eutrino \textbf{O}bservatory, an upcoming underground 
laboratory~\cite{VNAL, MKSI}. Although the natural abundance 
of $^{124}$Sn isotope is only $\sim$ 5.8\%, but its quite high Q$_{\beta\beta}$-value 
of 2287.7 keV makes it a good candidate for search of 0$\nu\beta\beta$~\cite{JDAW, VIVS}. 
A high Q$_{\beta\beta}$-value means that the search of 0$\nu\beta\beta$ process will 
be less affected by the natural radioactivity process and hence it increases the 
sensitivity factor of experiment~\cite{GGER, NEH3}. The High Energy Physics experimental 
group of Tata Institute of Fundamental Research (TIFR), Mumbai, has tested the cryogenic 
Sn bolometers (size of mg scale) and found these bolometers work very impressively 
with very good energy resolution at sub-Kelvin temperature~\cite{VNAN, VSIN}. The R\&D 
on approximately 1 kg $^{natural}$Sn prototype and the enrichment of $^{124}$Sn is in 
progress~\cite{VNAN}.\par

The sensitivity of an experiment can be decided by the following five 
important parameters; (1) Energy resolution ($\Delta$) at Q$_{\beta\beta}$, 
(2) Exposure ($\beta\beta_{isotope}$ mass$\times$time) ($\Sigma$), (3) Background rate ($\Lambda$), (4) Isotopic 
abundance (IA), and (5) Signal detection efficiency ($\epsilon_{expt}$). 
The smearing of 2$\nu\beta\beta$ events (B$_{2\nu}$) ($\tau^{2\nu}_{\frac{1}{2}}$ = 0.8-1.2$\times$10$^{21}$ yr)~\cite{VNAN} 
in the 0$\nu\beta\beta$ “Region of Interest” (ROI) is the irreducible background 
in the search of 0$\nu\beta\beta$. This can be minimized by using a detector with very 
good energy resolution. Therefore, the cryogenic bolometers will 
be a novel technique in the search of 0$\nu\beta\beta$ decay. \par

\section{Neutrino parameters and 0$\nu\beta\beta$ half-life}
In the simplest case, 0$\nu\beta\beta$ decay is mediated by the 
virtual exchange of a light Majorana neutrino in the absence of right-handed currents. 
The half-life (${\tau_{\frac{1}{2}}^{0\nu}}$) of 0$\nu\beta\beta$ isotopes can be 
expressed as~\cite{RGHR}

\begin{equation}
\Big [\tau^{0\nu}_{\frac{1}{2}}\Big]^{-1} ~ = ~ G^{'0 \nu} ~ g_{A}^{4} ~ | M^{0 \nu} |^{2} ~ \Bigg[\frac{\langle m_{\beta\beta}\rangle ^{2}}{m_{e}^{2}}\Bigg] 
                                      \equiv ~ G^{0 \nu} ~ | M^{0 \nu} |^{2} ~ \Bigg[\frac{\langle m_{\beta\beta}\rangle ^{2}}{m_{e}^{2}}\Bigg]. 
\label{eq:Core_Rel}
\end{equation}
\noindent
Here, G$^{'0 \nu}$ is the known phase space factor, G$^{0\nu}$ is the phase space 
factor combined with the weak axial vector coupling constant (g$_{A}$), $|M^{0\nu}|$ is the
Nuclear Physics matrix element and $m_{e}$ is the mass of the electron. To avoid the ambiguity of g$_{A}$ in the presence of nuclear medium, 
its free nucleon value (g$_{A}$ = 1.269)~\cite{JENG} is adopted. The effective Majorana 
neutrino mass is given by\cite{ADUE}: 
\begin{equation}
\langle m_{\beta\beta}\rangle ~ = ~ \big| ~ \sum^{j~=~0,\alpha,\beta}_{\gamma~=~1,2,3} ~~ e^{\gamma j}~|U_{e\gamma}|^{2}~~m_{\gamma} ~\big|, 
\label{eq:First_Mbb}
\end{equation}
\noindent
which depends on the neutrino masses (m$_{\gamma}$ for eigenstate $\nu_{\gamma}$), Majorana 
phases ($\alpha$, $\beta$) and PMNS (Pontecorvo-Maki-Nakagawa-Sakata) mixing matrix 
(U)\cite{RGHR, ADUE, GBEN}. Expansion of Eq. \hyperref[eq:First_Mbb]{4} will provide the $\langle$m$_{\beta\beta}$$\rangle$ as~\cite{ADUE}
\begin{equation}
\langle m_{\beta\beta}\rangle =  \big|c_{12}^{2}~c_{13}^{2}~m_{1}~+~s_{12}^{2}~c_{13}^{2}~m_{2}~e^{i\alpha}~+~s_{13}^{2}~m_{3}~e^{i(\beta-2\delta)} \big|.
\label{eq:Final_Mbb}
\end{equation} 
\noindent
The value of $\langle$m$_{\beta\beta}$$\rangle$ depends on sines (s) and cosines (c) 
of the leptonic mixing angles $\theta_{ij}$, the mass eigenvalues ($m_{\gamma}$), Majorana 
Phases e$^{i\alpha}$ = e$^{i\beta}$ = $\pm$1 and the CP violating phase e$^{-i2\delta}$ = 1. 
The measurement of mass-squared splitting ($\delta$m$^{2}_{\odot}$ 
= $\Delta$m$^{2}_{21}$ and $\Delta$m$^{2}_{atm}$ = $\frac{1}{2}$$|\Delta$m$^{2}_{31}$~+~$\Delta$m$^{2}_{32}|$) 
allows two hierarchy configurations for the mass eigenstates: either “Inverted 
Hierarchy” (IH) ($m_{3} <m_{1} <m_{2}$) or “Normal Hierarchy” (NH) ($m_{1} < m_{2} <m_{3}$) 
~\cite{WMAN, GBEN}. The allowed range for $\langle$m$_{\beta\beta}$$\rangle$ as a function 
of the lightest neutrino mass m$_{min}$ can be constrained by the experimental 
measurements of the neutrino mixing parameters. The lower and upper range of 
$\langle$m$_{\beta\beta}$$\rangle$ is derived from the cutoff choice m$_{min}$ = 10$^{-5}$ eV 
\begin{eqnarray}\nonumber
  IH: 1.765550\times10^{-2} ~(eV) \leq |\langle m_{\beta\beta}\rangle| \leq 4.981276\times10^{-2} ~(eV) \\
  NH: 1.363476\times10^{-3} ~(eV) \leq |\langle m_{\beta\beta}\rangle| \leq 4.093182\times10^{-3} ~(eV). 
\label{eq:hierarchy}
\end{eqnarray}
\par
The precise calculations of G$^{0\nu}$ and $|M^{0\nu}|$ are needed in order to 
translate the experimental values of the 0$\nu\beta\beta$ half-lives into 
$\langle$m$_{\beta\beta}$$\rangle$. With an uncertainty of approximately 7 \%, 
G$^{0\nu}$ is well known~\cite{PGUO}. On the other hand, the calculation of
$|M^{0\nu}|$ is a difficult task involving the details of the underlying theoretical 
models. Several different theoretical models have been used to compute $|M^{0\nu}|$ 
for the different A$_{\beta\beta}$ such as interacting shell model (ISM)~\cite{MJNE}, 
quasiparticle random phase approximation (QRPA) (and its variants)~\cite{FEDO, 
DLFG}, interacting boson model (IBM-2)~\cite{JBAR}, angular momentum projected 
hartree-fock bogoliubov method (PHFB)~\cite{PKRT}, generating coordinate 
method (GCM) and energy density functional method (EDF)~\cite{JENG, RRTO}. 
Deviations among their results are the main sources of theoretical uncertainties 
in the required sensitivity. \par 
\setlength{\tabcolsep}{1.55em}
\begin{table}[h!]
\small
\captionsetup{font=small}
\def\arraystretch{1.7}
\caption{Nuclear matrix elements for $^{124}$Sn isotope extracted from the references ~\cite{VNAN, RGHR, JKFI}.}
\label{table:NME}
\centering
\begin{center}
\begin{tabular}{lc}
\hline \hline 
 {Theoretical Model (Scheme)}                       & {$|M^{0\nu}|$}               \\ \hline 
 
 {Projected Hartree-Fock-Bouglebov (PHFB)}           & 6.04  \\ 
 {Generating coordinate method (GCM)}                & 4.81  \\ 
 {Interacting boson model (IBM)}                     & 3.53  \\ 
 {Shell Model (SM)}                                  & 2.62  \\ \hline \hline
  
\end{tabular} \end{center}
\end{table}
For $^{124}$Sn isotope, $|M^{0\nu}|$ along with the corresponding theoretical 
models are listed in Table \hyperref[table:NME]{1}. In the given range 
of $|M^{0\nu}|$, the PHFB and SM are in most optimistic and most conservative 
scenario, respectively. Therefore, the required sensitivity corresponding to 
other $|M^{0\nu}|$ will lie in between this range. Using the range of $\langle$m$_{\beta\beta}$$\rangle$ 
from Eq. \hyperref[eq:hierarchy]{6} and $|M^{0\nu}|$ from  Table \hyperref[table:NME]{1}, with the 
help of Eq. \hyperref[eq:Core_Rel]{3}, corresponding benchmark sensitivities can be 
calculated in terms of ${\tau_{\frac{1}{2}}^{0\nu}}$. The value of combined function ($F_{n}$) for 
PHFB and SM models are adopted from Ref.~\cite{VNAN} 
\begin{eqnarray}\nonumber
  F_{n} = G^{0\nu}.|M^{0\nu}|^{2} =8.569\times10^{-13} yr^{-1} (PHFB) \\
  ~~~~~~~~~~~~~~~~~~~~~=1.382\times10^{-13} yr^{-1} (SM).~~~~~
\label{eq:NME}
\end{eqnarray}
Using Eqns. \hyperref[eq:Core_Rel]{3}, \hyperref[eq:hierarchy]{6} and \hyperref[eq:NME]{7}, 
the required sensitivities in the form of ${\tau_{\frac{1}{2}}^{0\nu}}$ are
\begin{eqnarray}\nonumber
  PHFB \equiv IH:~ 1.228091\times10^{26} (yr) < \tau_{\frac{1}{2}}^{0\nu} < 9.776263\times10^{26} (yr)\\\nonumber
  ~~~~~~~~~~~~~NH:~ 1.818819\times10^{28} (yr) < \tau_{\frac{1}{2}}^{0\nu} < 1.639143\times10^{29} (yr) \\\nonumber
  SM \equiv IH:~ 7.614697\times10^{26} (yr) < \tau_{\frac{1}{2}}^{0\nu} < 6.061708\times10^{27} (yr) \\
  ~~~~~~~~NH:~ 1.127747\times10^{29} (yr) < \tau_{\frac{1}{2}}^{0\nu} < 1.016340\times10^{30} (yr).
\label{eq:Hierarchy_HF}
\end{eqnarray}

The current generation of oscillation experiments may reveal Nature’s choice 
among the two hierarchy options. Moreover, the combined cosmology data may 
provide a measurement on the sum of $m_{i}$~\cite{FCAP, AGIU}. Thus, 
it can be expected that the ranges of parameter space in 0$\nu\beta\beta$ 
searches will be further constrained. \par

From the experimental point of view, the measurement of 
half-life ${\tau_{\frac{1}{2}}^{0\nu}}$ of 0$\nu\beta\beta$ relies just on 
the observed signal ($S_{0\nu}$ (0$\nu\beta\beta$-events)). The relationship 
between ${\tau_{\frac{1}{2}}^{0\nu}}$ and observed $S_{0\nu}$ can be derived 
from the law of radioactive decay 
\begin{equation}
\Big[ \tau^{0\nu}_{\frac{1}{2}}\Big]^{-1} ~ = ~ \big[ {\rm log_e 2}\big]^{-1} ~ \bigg[ \frac{A}{N_{A}} \bigg] ~ \bigg[ \frac{1}{\Sigma} \bigg]  ~ \bigg[ \frac{S_{0\nu}}{\varepsilon_{ROI}} \bigg],  
\label{eq:Formula}
\end{equation}
\noindent
where A is the molar mass of the source $A_{\beta\beta}$, $N_{A}$ is the 
Avogadro Number and $\varepsilon_{ROI}$ is the efficiency of selected ROI. \par

In the search of 0$\nu\beta\beta$ decay, the ROI around the Q$_{\beta\beta}$ value 
could be symmetric and asymmetric. The symmetric FWHM ROI at Q$_{\beta\beta}$ value 
is most often choice of experiments. The ROI in the current study is taken to be 
the FWHM window centered at Q$_{\beta\beta}$, such that the efficiency $\varepsilon_{ROI}$~=~76.1~\%. 
Every experiment needs to use an enriched isotope for obtaining 
the better sensitivity. Therefore for simplicity and being easily 
convertible, both the IA of the 0$\nu\beta\beta$ isotopes in the 
target and the other experimental efficiencies ($\varepsilon_{expt}$) 
are taken to be~100~\%. In practice, the required combined 
exposure $\Sigma^{'}$ of $A_{\beta\beta}$, can be converted from the ideal 
$\Sigma$ of the present work via $\Sigma^{'}$ = $\Sigma$/(IA. $\varepsilon_{expt}$).

\section{Experimental constraints on sensitivity}

The background events are always present in realistic experiments which degrade 
the sensitivities of the identifying spectral peaks at Q$_{\beta\beta}$. 
The source of background in the search of 0$\nu\beta\beta$ can be divided
into two categories: intrinsic and ambient. The ambient background is mostly 
induced by external $\gamma$-rays, especially from trace radioactivity present in the experimental 
hardware and cosmogenically activated isotopes in the vicinity of target volume. 
The total ambient background counts N$_{a}$ in the 0$\nu\beta\beta$ ROI can be obtained 
from the following expression
\begin{equation}
N_{a}~ =~\Lambda_{a}.~\Sigma.~[\Delta.~Q_{\beta\beta}],
\label{eq:Ambient}
\end{equation}
where $\Lambda_{a}$ is the flat ambient background rate in the units of counts/tonne-year-keV (/tyk). 
The intrinsic background in the 0$\nu\beta\beta$ search come from the 2$\nu\beta\beta$ 
decay process. It is therefore inherently associated with the A$_{\beta\beta}$ 
and directly proportional to $\Sigma$. The finite detector resolution leads the 
irreducible B$_{2\nu}$ events which contaminates the 0$\nu\beta\beta$ ROI. 
Therefore, the sum (B$_{0}$ = B$_{2\nu}$+N$_{a}$) would be the total background counts in 
the selected ROI.  \par

\begin{figure*}
  \centering
  \begin{minipage}[t]{0.48\textwidth}
    \includegraphics[width=\textwidth]{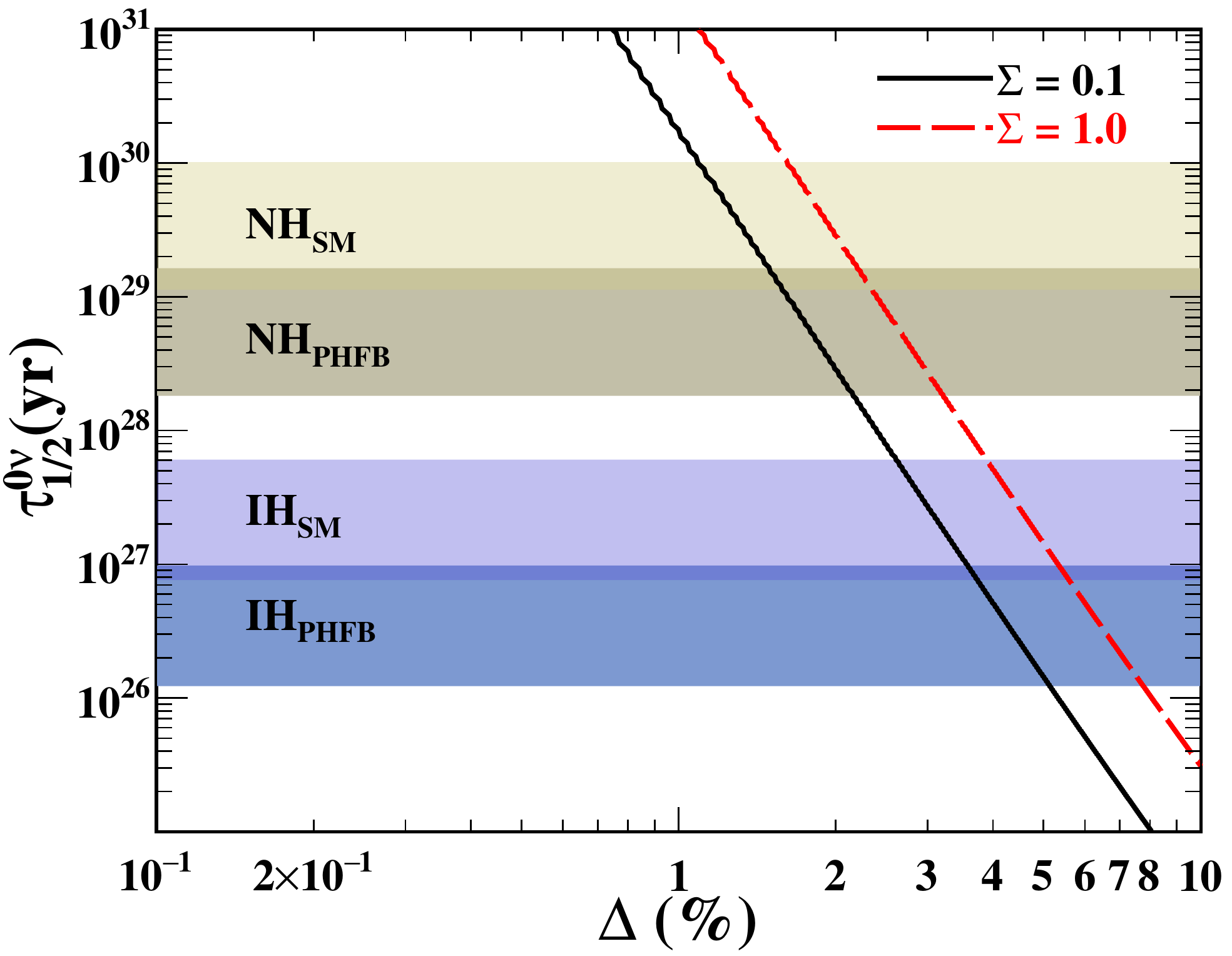}
    \caption{Variation of 0$\nu\beta\beta$ decay half-life with $\Delta$. 
The contamination of B$_{2\nu}$ events in ROI is shown by continuous and dotted line for $\Sigma$ = 0.1
and 1.0 ty, respectively. To get the expectations of B$_{2\nu}$ events in the maximum range of 
$|M^{0\nu}|$ uncertainty, the IH and NH bands are also superimposed.}
    \label{fig:Half_life}
  \end{minipage}
  \hfill
  \begin{minipage}[t]{0.48\textwidth}
    \includegraphics[width=\textwidth]{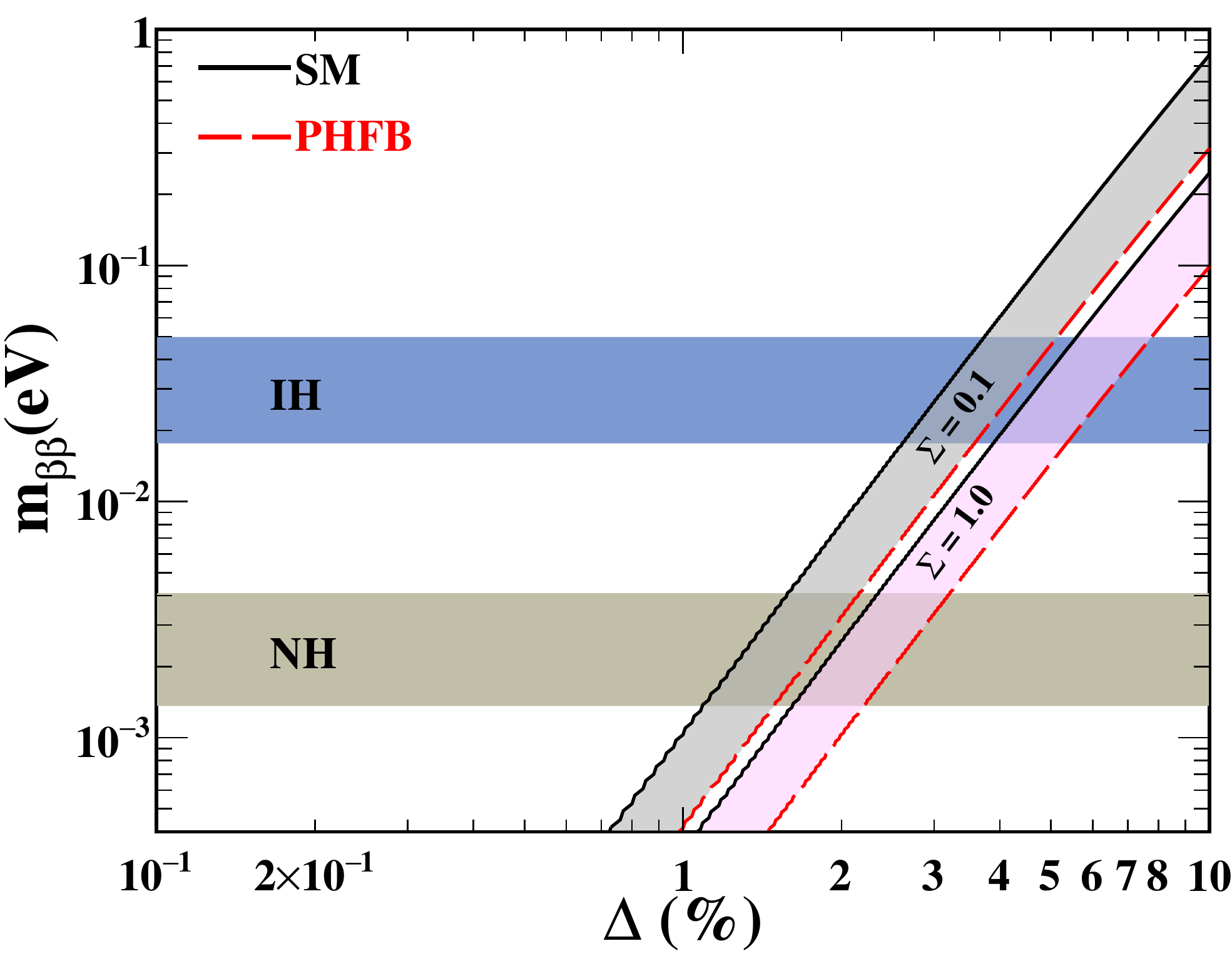}
    \caption{Variation of 
$\langle$m$_{\beta\beta}$$\rangle$ with $\Delta$. The uncertainty of $|M^{0\nu}|$, leads the uncertainty 
in $\langle$m$_{\beta\beta}$$\rangle$ due to the B$_{2\nu}$ events, which is shown in the form 
of the band (other than the Hierarchy bands).}
    \label{fig:Mbb}
  \end{minipage}
  \hfill
\end{figure*}

If the ambient background reduces to a minimum level (N$_{a}$ = 0), the irreducible 
background B$_{2\nu}$ would remain in the ROI. The contamination of B$_{2\nu}$ mainly depends 
on the detector $\Delta$ and $\Sigma$. The lower limit on $\tau^{0\nu}_{\frac{1}{2}}$ due to 
only B$_{2\nu}$ as a function of $\Delta$ for $\Sigma$ = 0.1 and 1.0 tonne-year (ty) are represented by the 
continuous and dotted lines respectively in Fig. \hyperref[fig:Half_life]{2}. The IH and NH 
bands corresponding to the SM and PHFB $|M^{0\nu}|$ are also superimposed (From Eqns. \hyperref[eq:hierarchy]{6} 
and \hyperref[eq:Hierarchy_HF]{8}) to get the prospects of B$_{2\nu}$ for future $^{124}$Sn isotope 
based experiments. \par

The conversion of $\tau^{0\nu}_{\frac{1}{2}}$ in $\langle$m$_{\beta\beta}$$\rangle$ sensitivity face 
the theoretical uncertainty of $|M^{0\nu}|$ (Eqn. \hyperref[eq:Core_Rel]{3}). Therefore, the $\langle$m$_{\beta\beta}$$\rangle$ 
sensitivity due to B$_{2\nu}$ form band structure (apart from the IH and NH bands) in the 
$\langle$m$_{\beta\beta}$$\rangle$~vs~$\Delta$ parameter space as shown in Fig. \hyperref[fig:Mbb]{3}. 
The upper and lower line of $|M^{0\nu}|$ uncertainty band arises due to the $|M^{0\nu}|$ of SM and PHFB, 
respectively. This leads that the $|M^{0\nu}|$ of SM would impose severe requirements on experimental 
sensitivity in comparison to the PHFB.   \par
With the maximum range of $|M^{0\nu}|$ uncertainty for $\Sigma_{0}$ = 1.0 ty to cover the NH, the safe zone 
from B$_{2\nu}$ events begins at $\Delta$ $<$ 1.61\% for SM and $\Delta$ $<$ 2.19\% for PHFB. 
The safe zone for IH case begins from $\Delta$ $<$~3.88\% for SM and $\Delta$ $<$~5.34\% 
for PHFB. This leads that the TIN.TIN experiment would be very less affected by the 
B$_{2\nu}$ events ($\sim$ 3.08 $\times10^{-6}$ counts) if it reach to the energy 
resolution $\Delta_{0}$ = 0.5\% at Q$_{\beta\beta}$, which is close to the achieved energy resolution~=~0.31\% at Q$_{\beta\beta}$ 
of the CUORE experiment (Bolometric detector using $^{130}$Te)~\cite{ALDU}.

\section{Statistical significance of signal}
Rare event physics search like 0$\nu\beta\beta$ and dark matter naturally demands 
the very low background experiment~\cite{NEH1}. The understanding of background and its 
suppression would significantly improves the experimental sensitivity. In the design stage of 
experiments, the averaged N$_{a}$ and B$_{2\nu}$ can be precisely estimated from the 
prior knowledge of the most relevant sources of background and simulation studies~\cite{NEH2, VSNG, NEH3, NEH4}. 
The low background counts in the ROI are subjected to the Poisson fluctuation. Excess 
of counts from expected background may originate from the upward fluctuations of the 
background channels. The discovery potential (D.P.) and sensitivity level (S.L.) can 
be expressed in the frame of background fluctuation. In order to get the strong evidence, 
we have calculated the signal counts with 3$\sigma$ S.L. and 5$\sigma$ is expressed 
for D.P. \par

\begin{figure} 
  \centering
  \includegraphics[width=8.5cm]{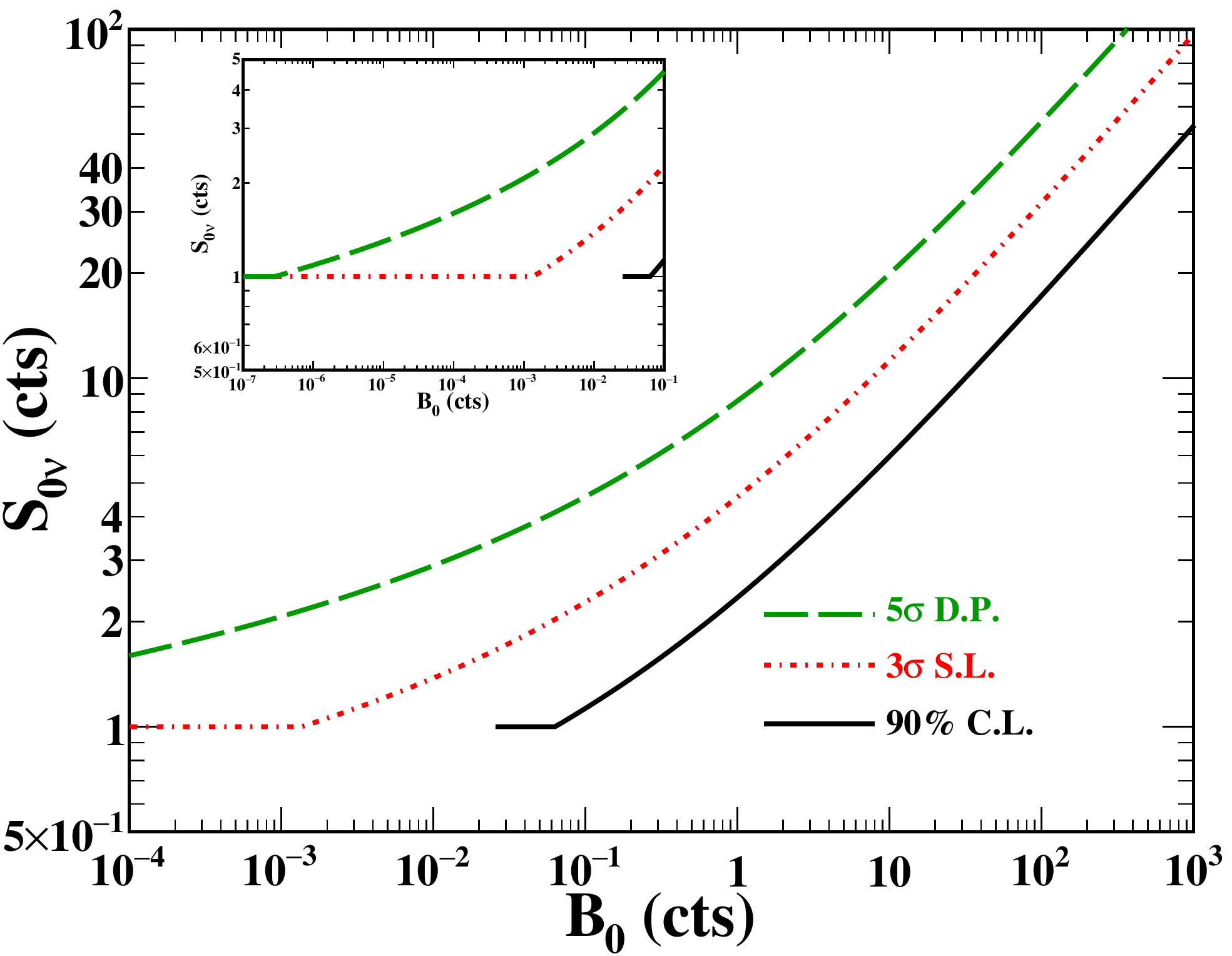}
  \caption{Variation of S$_{0\nu}$ corresponding to B$_{0}$ under the 
3$\sigma$ S.L., 5$\sigma$ D.P. and at 90\% C.L. schemes of signal identification.}
  \label{fig:S_vs_B}
\end{figure}

The Poisson distribution is discrete and provide the significance level for certain values only.
The continuous representation of the Poisson distribution is obtained by normalized upper incomplete 
gamma function and it gives the probability distribution~\cite{SHAB}

\begin{equation}
F(k) = \frac{\Gamma(k+1, \lambda)}{\Gamma(k+1)}, k>0, with 
\label{eq:GAMMA}
\end{equation}

\begin{equation}
\Gamma(k, \lambda) = \int_{\lambda}^{\infty} e^{-t}~t^{k-1}~dt ~~~~ and ~~~~ \Gamma(k) = \int_{0}^{\infty} e^{-t}~t^{k-1}~dt,
\label{eq:_Final_Gamma}
\end{equation}

\noindent
where $\lambda$ is the mean value of distribution, k is the number of counts, $\Gamma(k, \lambda)$ is the upper 
incomplete gamma function and $\Gamma(k)$ is the ordinary gamma function. The variation in 
sensitivity became free from the discrete steps (with Eq.~\hyperref[eq:GAMMA]{11}). For completeness, the 
signal counts at 90\% C.L. are also calculated from the Poisson distribution as illustrated in 
Fig. \hyperref[fig:S_vs_B]{4}. \par

For very low expected background, the requirement of S$_{0\nu}$ for an experiment is chosen to be a 1 event. 
This leads to the same sensitivity at the background free level. The background free criteria depend 
on the chosen statistical scheme. The background free scenario is shown in Fig. \hyperref[fig:S_vs_B]{4} 
from the horizontal line limiting at S$_{0\nu}$ = 1 event. As the background decreases the significance of 
S$_{0\nu}$ increases. This increment in significance is shown in Fig. \hyperref[fig:S_vs_B]{4} by flattened line.
On reaching the background free criteria, the extension of 90\% C.L. is extended up to the 2$\sigma$ level while 
the 3$\sigma$ S.L. is extended up to the 5$\sigma$ D.P. \par

After using these two schemes for the identification of S$_{0\nu}$, 
the ${\tau_{\frac{1}{2}}^{0\nu}}$ sensitivity of~Eq.~\hyperref[eq:Formula]{9} would take the following form
\begin{equation}
\Big[\tau^{0\nu}_{\frac{1}{2}}\Big]^{-1} ~ = ~ \big[{\rm log_e 2}\big]^{-1} ~ \bigg[\frac{A}{N_{A}} \bigg] ~ \bigg[\frac{1}{\Sigma} \bigg]  ~ \bigg[\frac{S_{90\%}~|~S_{3\sigma}~|~S_{5\sigma}}{\varepsilon_{ROI}} \bigg],
\label{eq:Final_Formula}
\end{equation}
where S$_{0\nu}$ of~Eq. \hyperref[eq:Formula]{9} is replaced by S$_{90\%}$, S$_{3\sigma}$ and S$_{5\sigma}$ to 
obtain the ${\tau_{\frac{1}{2}}^{0\nu}}$ sensitivity at 90\% CL, 3$\sigma$ S.L. and 5$\sigma$~D.P. level, respectively. 
Under these two schemes the required sensitivity for $^{124}$Sn isotope is studied in terms of 
required $\Lambda$, $\Sigma$ at the $\Delta_{0}$~=~0.5\% at Q$_{\beta\beta}$. These 
sensitivities are calculated with the aim to reach the most conservative (min.) and most optimistic 
(max.) regime of IH and NH (see Eq. \hyperref[eq:hierarchy]{6}).

\section{0$\nu\beta\beta$ half-life sensitivity as a function 
of $\Sigma$ and $\Lambda$ at $\Delta_{0}$}

\begin{figure*}
  \centering
  \begin{minipage}[t]{0.48\textwidth}
    \includegraphics[width=\textwidth]{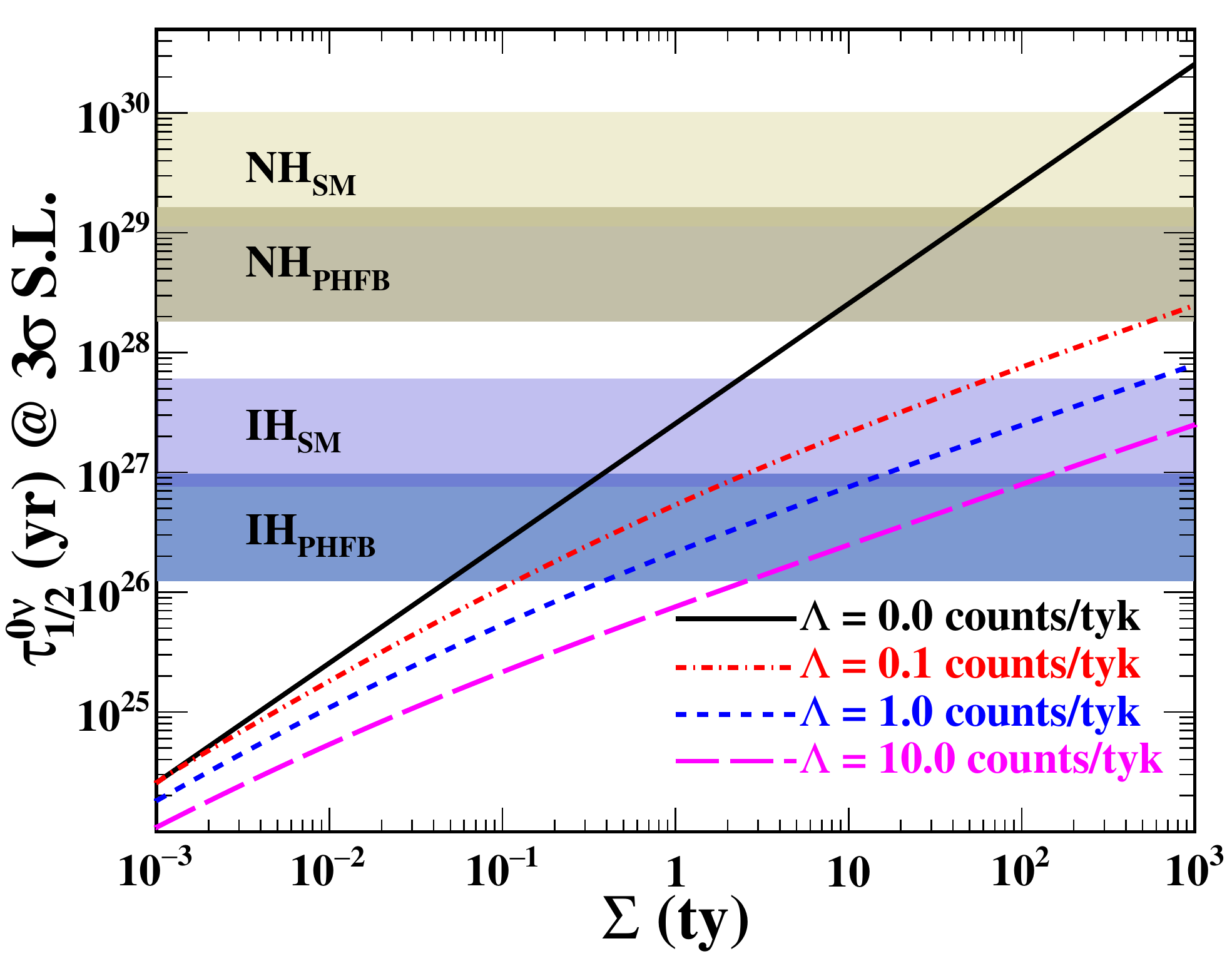}
    \caption{Signal identification at 3$\sigma$ S.L. in ${\tau_{\frac{1}{2}}^{0\nu}}$ versus $\Sigma$ at $\Delta_{0}$ 
for $\Lambda$~=~(0, 0.1, 1.0, 10.0)/tyk. The IH and NH bands are superimposed for 
both PHFB and SM $|M^{0\nu}|$ models.}
    \label{fig:3Sigma_SL}
  \end{minipage}
  \hfill
  \begin{minipage}[t]{0.48\textwidth}
    \includegraphics[width=\textwidth]{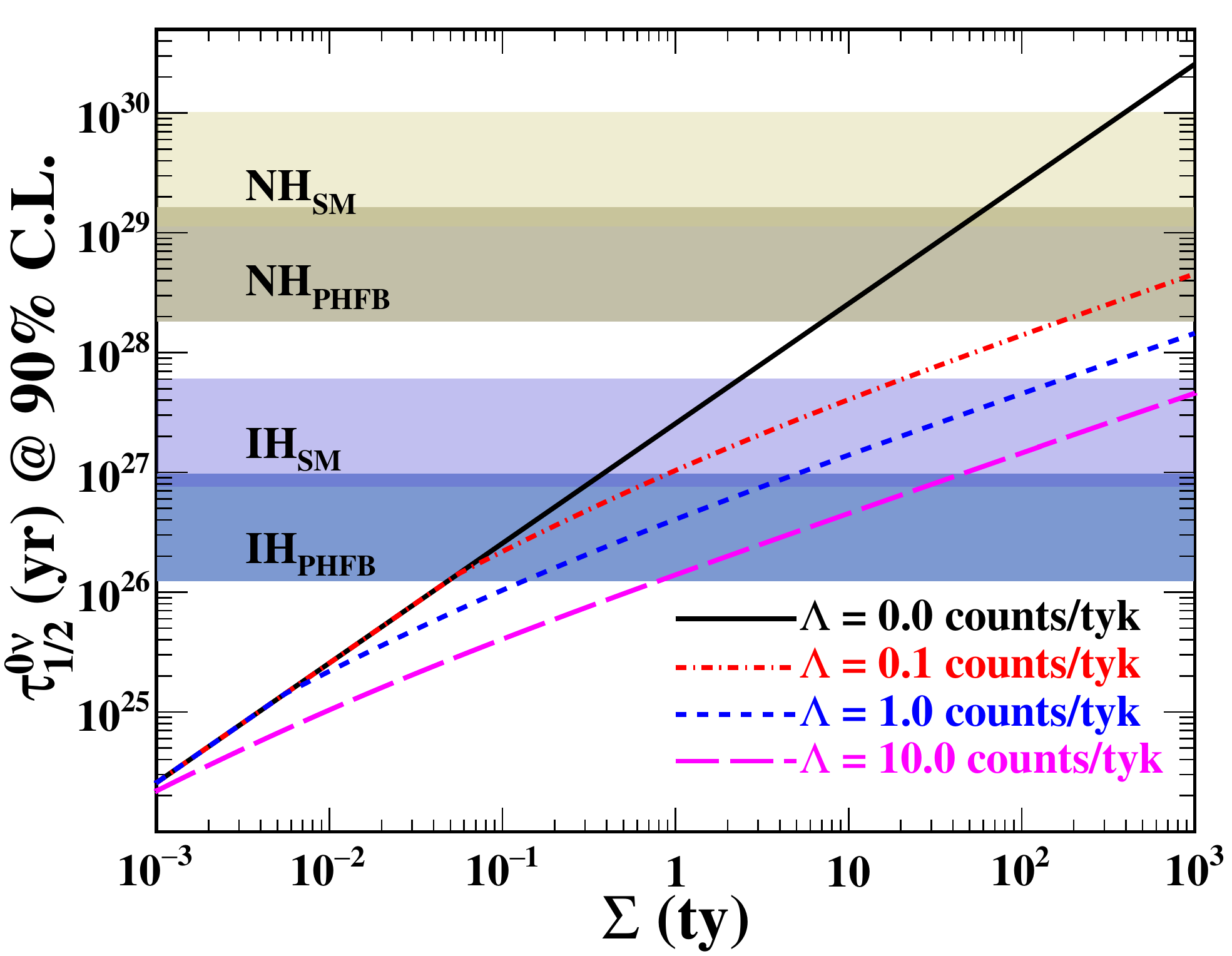}
    \caption{Signal identification at 90\% C.L. in ${\tau_{\frac{1}{2}}^{0\nu}}$ versus $\Sigma$ at $\Delta_{0}$ 
for $\Lambda$~=~(0, 0.1, 1.0, 10.0)/tyk. The IH and NH bands are superimposed for 
both PHFB and SM $|M^{0\nu}|$ models.}
    \label{fig:90_CL}
  \end{minipage}
  \hfill
\end{figure*}

The accessible physics with 0$\nu\beta\beta$ experiments is the effective mass of 
Majorana neutrinos $\langle$m$_{\beta\beta}$$\rangle$, which is the linear combination of neutrino 
mass eigenstates. Therefore, the minimum desired experimental sensitivity of the TIN.TIN experiment is to 
probe the IH mass region. The $\tau^{0\nu}_{\frac{1}{2}}$ is 
inversely proportional to the $\langle$m$_{\beta\beta}$$\rangle$. The variation of 
$\tau^{0\nu}_{\frac{1}{2}}$ at 3$\sigma$ S.L. and 90\% C.L. as a function of $\Sigma$ at a fixed 
$\Delta_{0}$ with various background rates ($\Lambda$) is depicted 
in Figs. \hyperref[fig:3Sigma_SL]{5} and \hyperref[fig:90_CL]{6} respectively. The hierarchy bands arises from uncertainty of $|M^{0\nu}|$ 
and range of $\langle$m$_{\beta\beta}$$\rangle$ (Eqns. \hyperref[eq:hierarchy]{6} and \hyperref[eq:Hierarchy_HF]{8}) 
is also superimposed over it. \par

The required sensitivity in terms of exposure $\Sigma$ and benchmark background rate  
$\Lambda$~=~(0, 0.1, 1.0, 10.0)/tyk, at $\Delta_{0}$, to just enter the hierarchy 
is summarized in Table \hyperref[table:OPTIMISTIC]{2}. In order to enter the IH$_{PHFB}$ mass region with 
$\Lambda$ = 0.1/tyk the TIN.TIN experiment must have $\Sigma$ = 0.12 ty at 3$\sigma$ S.L. 
($\Sigma$ = 4.80$\times10^{-2}$ ty at 90\% C.L.), but the IH$_{SM}$ mass region requires 
$\Sigma$ = 1.74 ty at 3$\sigma$ S.L. ($\Sigma$ = 0.62 ty at 90\% C.L.). Similarly, having the 
same background rate of $\Lambda$ = 0.1/tyk requires $\Sigma$ = 5.45$\times10^{2}$ ty at 3$\sigma$ S.L. 
(1.67$\times10^{2}$ ty at 90\% C.L.) for NH$_{PHFB}$ and 2.01$\times10^{4}$ ty at 3$\sigma$ S.L. 
(6.04$\times10^{3}$ ty at 90\% C.L.) for NH$_{SM}$. The uncertainty of $|M^{0\nu}|$ leads the uncertainty in required 
sensitivity. Therefore, precise calculation of $|M^{0\nu}|$ from different model is the main requirement. \par

\setlength{\tabcolsep}{0.15em}
\begin{table}[h!]
\small
\def\arraystretch{1.5}
\captionsetup{font=small}
\caption{The required $\Sigma$ corresponding to the benchmark $\Lambda$~=~(0, 0.1, 1.0, 10.0)/tyk 
in the most optimistic scenario to just enter the hierarchy.}
\label{table:OPTIMISTIC}
\centering
\begin{center}
\begin{tabular}{ccccc|cccc}
\hline \hline 
\multicolumn{1}{c|}{}& \multicolumn{4}{c|}{\bf{${\big[\tau_{\frac{1}{2}}^{0\nu}\big]}$$^{IH}_{min}$}} & \multicolumn{4}{c}{\bf{${\big[\tau_{\frac{1}{2}}^{0\nu}\big]}$$^{NH}_{min}$}}  \\ \cline{2-9}
\multicolumn{1}{c|}{}& \multicolumn{8}{c}{\bf{SM}}  \\ \cline{1-9} 

\multicolumn{1}{l|}{\bf{$\Lambda$ (/tyk)}}             &\bf{0.0} &\bf{0.1} &\bf{1.0}  &\bf{10.0} &\bf{0.0} &\bf{0.1}            &\bf{1.0}            &\bf{10.0}\\ \cline{1-9} 
\multicolumn{1}{l|}{\bf{$\Sigma$ (ty)} @ 3$\sigma$ S.L.} &0.30     &1.74     &10.15     &92.12    &44.03  &2.01$\times$10$^{4}$ &2.18$\times$10$^{5}$ &2.47$\times10^{6}$\\\cline{1-9} 
\multicolumn{1}{l|}{\bf{$\Sigma$ (ty)} @ 90\% C.L.}    &0.30      &0.62     &3.18     &27.84     &44.03  &6.04$\times$10$^{3}$ &6.28$\times$10$^{4}$ &7.04$\times10^{5}$\\\cline{1-9} 

\multicolumn{1}{l}{}  &  \multicolumn{8}{c}{\bf{PHFB}} \\ \cline{1-9} 
\multicolumn{1}{l|}{\bf{$\Sigma$ (ty)} @ 3$\sigma$ S.L.}&4.80$\times10^{-2}$ &0.12  &0.38  &2.55    &7.11   &5.45$\times$10$^{2}$ &5.27$\times10^{3}$ &5.82$\times10^{4}$\\ \cline{1-9}
\multicolumn{1}{l|}{\bf{$\Sigma$ (ty)} @ 90\% C.L.}  &4.80$\times10^{-2}$&4.80$\times10^{-2}$&0.13&0.78  &7.11 &1.67$\times10^{2}$ &1.56$\times10^{3}$ &1.67$\times10^{4}$ \\ \cline{1-9} \hline \hline   

\end{tabular}
\end{center}
\end{table}

\begin{figure*}
  \centering
  \begin{minipage}[t]{0.48\textwidth}
    \includegraphics[width=\textwidth]{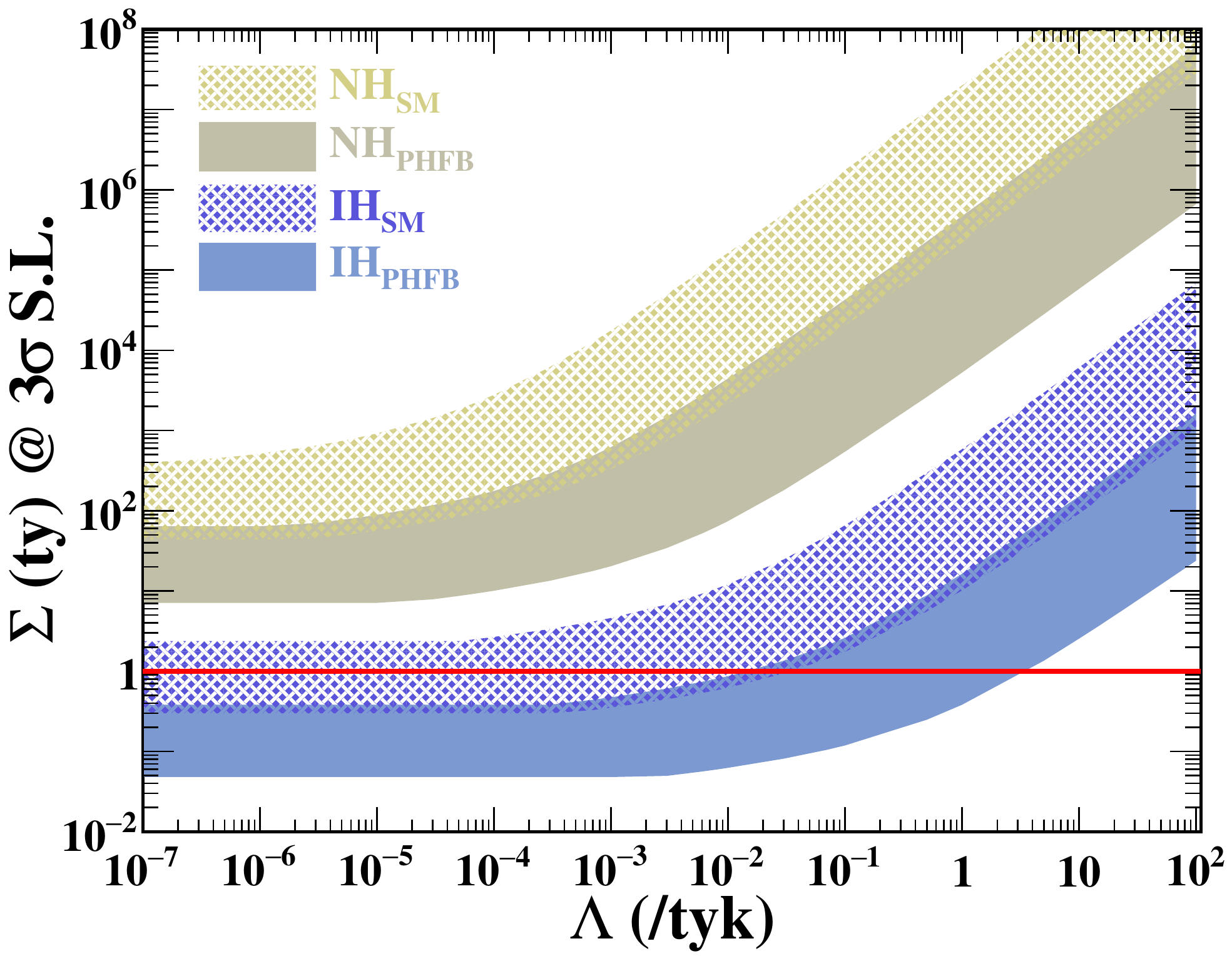}
    \caption{Signal identification to cover completely and to just enter the hierarchy 
with 3$\sigma$ S.L. in $\Sigma$ versus $\Lambda$ space for $^{124}$Sn at $\Delta_{0}$, in complete regime (min. to max.) of 
$\langle$m$_{\beta\beta}$$\rangle$ for IH and NH, using the $M^{0\nu}$ of SM and PHFB models.}
    \label{fig:3Sigma_SL_EXPO}
  \end{minipage}
  \hfill
  \begin{minipage}[t]{0.48\textwidth}
    \includegraphics[width=\textwidth]{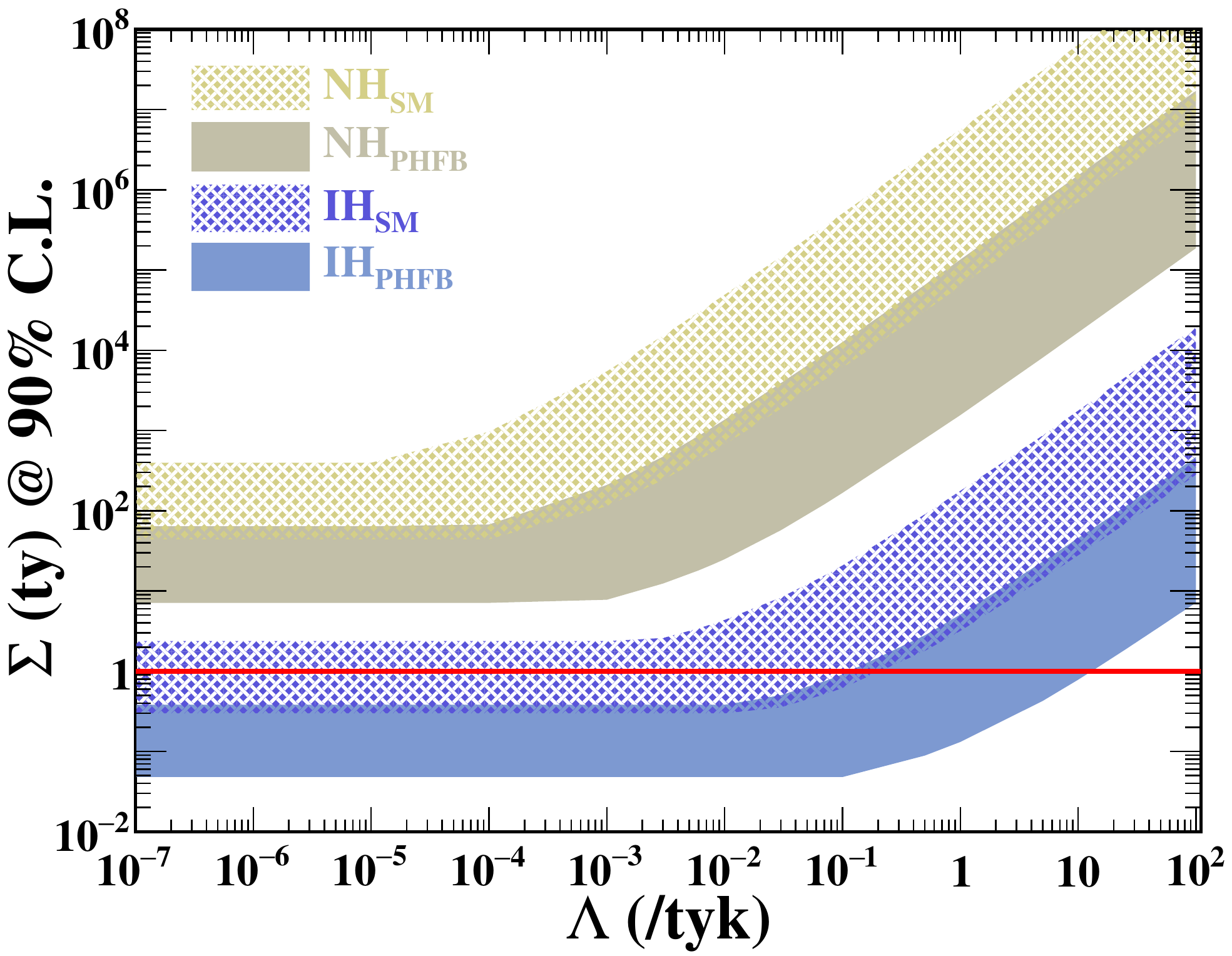}
    \caption{Signal identification to cover completely and to just enter the hierarchy 
with 90\% C.L. in $\Sigma$ versus $\Lambda$ space for $^{124}$Sn at $\Delta_{0}$, in complete regime (min. to max.) of 
$\langle$m$_{\beta\beta}$$\rangle$ for IH and NH, using the $M^{0\nu}$ of SM and PHFB models.}
    \label{fig:90_CL_EXPO}
  \end{minipage}
  \hfill
\end{figure*}

The potential of improvement in background is explained in the parameter space of $\Sigma$ and $\Lambda$ 
at $\Delta_{0}$ in conjunction with the uncertainty bands of $|M^{0\nu}|$ for both the IH and NH 
(Figs. \hyperref[fig:3Sigma_SL_EXPO]{7} and \hyperref[fig:90_CL_EXPO]{8}). The reduction in the background leads the controllable requirement imposed on $\Sigma$. 
Therefore, the background improvement is a necessity for the experiment. This improvement in the background plays 
crucial role in order to cover the hierarchy region completely. The required sensitivity in 
terms of $\Sigma$ at $\Delta_{0}$ to completely cover both the hierarchy are summarized in 
Table \hyperref[table:CONSERVATIVE]{3} for both $|M^{0\nu}|$ at 3$\sigma$ S.L. and 90\% C.L. \par

At the earlier chosen background rate $\Lambda$ = 0.1/tyk, the coverage of IH$_{PHFB}$ requires $\Sigma$ = 2.61 ty 
at 3$\sigma$ S.L. ($\Sigma$ = 0.91 ty at 90\% C.L.) while for IH$_{SM}$ this requirement becomes $\Sigma$ = 65.88 ty 
at 3$\sigma$ S.L. ($\Sigma$ = 20.82 ty at 90\% C.L.). Similarly, in order to cover the NH$_{PHFB}$ at 
3$\sigma$ S.L. requires $\Sigma$ = 4.27$\times10^{4}$ ($\Sigma$ = 1.27$\times10^{4}$ ty at 90\% C.L.) and coverage 
of NH$_{SM}$ demanded an exposure of $\Sigma$ = 1.76$\times10^{6}$ ty at 3$\sigma$ S.L. ($\Sigma$ = 5.08$\times10^{5}$ ty 
at 90\% C.L.).  \par

\setlength{\tabcolsep}{0.15em}
\begin{table}[h!]
\small
\def\arraystretch{1.5}
\captionsetup{font=small}
\caption{The required sensitivity in terms of $\Sigma$ corresponding to the benchmark 
$\Lambda$~=~(0, 0.1, 1.0, 10.0)/tyk, for covering completely the hierarchy (most 
conservative scenario).}
\label{table:CONSERVATIVE}
\centering
\begin{center}
\begin{tabular}{ccccc|cccc}
\hline \hline 
\multicolumn{1}{c|}{}  & \multicolumn{4}{c|}{\bf{${\big[\tau_{\frac{1}{2}}^{0\nu}\big]}$$^{IH}_{max}$}} & \multicolumn{4}{c}{\bf{${\big[\tau_{\frac{1}{2}}^{0\nu}\big]}$$^{NH}_{max}$}}  \\ \cline{2-9}
\multicolumn{1}{c|}{}  & \multicolumn{8}{c}{\bf{SM}}  \\ \cline{1-9} 

\multicolumn{1}{l|}{\bf{$\Lambda$ (/tyk)}}             &\bf{0.0} &\bf{0.1} &\bf{1.0} &\bf{10.0} &\bf{0.0} &\bf{0.1}            &\bf{1.0}            &\bf{10.0}\\ \cline{1-9} 
\multicolumn{1}{l|}{\bf{$\Sigma$ (ty)} @ 3$\sigma$ S.L.} &2.37    &65.88     &5.86$\times10^{2}$   &6.11$\times10^{3}$  &3.97$\times10^{2}$  &1.76$\times$10$^{6}$ &2.00$\times$10$^{7}$ &2.28$\times10^{8}$\\\cline{1-9} 
\multicolumn{1}{l|}{\bf{$\Sigma$ (ty)} @ 90\% C.L.}    &2.37    &20.82    &1.77$\times10^{2}$   &1.77$\times10^{3}$   &3.97$\times10^{2}$ &5.08$\times$10$^{5}$ &5.69$\times$10$^{6}$ &6.42$\times10^{7}$\\\cline{1-9} 

\multicolumn{1}{l}{}  &  \multicolumn{8}{c}{\bf{PHFB}} \\ \cline{1-9} 
\multicolumn{1}{l|}{\bf{$\Sigma$ (ty)} @ 3$\sigma$ S.L.} &0.38  &2.61 &16.36 &1.51$\times10^{2}$  &64.05 &4.27$\times$10$^{4}$ &4.70$\times10^{5}$ &5.34$\times10^{6}$\\ \cline{1-9}
\multicolumn{1}{l|}{\bf{$\Sigma$ (ty)} @ 90\% C.L.}    &0.38  &0.91 &5.09  &45.68              &64.05  &1.27$\times10^{4}$ &1.35$\times10^{5}$ &1.52$\times10^{6}$ \\ \cline{1-9} \hline \hline   

\end{tabular}
\end{center}
\end{table}

The value of minimum exposure $\Sigma_{min}$ corresponding to 1 S$_{0\nu}$ event is obtained 
at very low background (close to $\Lambda$ = 0/tyk) and 
shown by the left flattened region in Figs. \hyperref[fig:3Sigma_SL_EXPO]{7} and \hyperref[fig:90_CL_EXPO]{8}. $\Sigma_{min}$ is an 
important parameter where each related experiment wants to reach by making improvement 
in the achieved background rate $\Lambda_{0}$~$\rightarrow$~$\Lambda$~=~0/tyk. The value of $\Sigma_{min}$ gives clear 
indication about the enhancement of required sensitivity in terms of $\Sigma$ with 
$\Lambda$ in the experiment. It has explicitly comes out that in order to  
just enter the IH$_{PHFB}$, it needs $\Sigma_{min}$ = 4.80$\times10^{-2}$ ty and for IH$_{SM}$ 
the value of $\Sigma_{min}$ = 0.30 ty. Similarly to enter the NH$_{PHFB}$ requires $\Sigma_{min}$~=~7.11 ty 
and for NH$_{SM}$ this value became 44.03 ty. In order to cover the IH$_{PHFB}$ requires 
$\Sigma_{min}$ = 0.38 ty and the coverage of IH$_{SM}$ requires 2.37 ty. The coverage of NH$_{PHFB}$ 
demands $\Sigma_{min}$ = 64.05 ty and for NH$_{SM}$ this requirement reaches up to 3.97$\times10^{2}$ ty. \par

\section{Summary and prospects}

The next generation neutrinoless double-beta decay experiments like TIN.TIN 
have a primary aim to probe the IH region. We have investigated the experimental 
parameters such as energy resolution, exposure and background rate to meet this goal in reference 
of background fluctuation sensitivity at 3$\sigma$ S.L. and 90\% C.L. This background fluctuation 
sensitivity study can be straightforward extended to the discovery potential for any experiment. \par

Our present study shows that the energy resolution of 0.5\% at Q$_{\beta\beta}$ for TIN.TIN detector is good enough to 
overcome the two neutrino double-beta decay background events in perspective to probe the IH. In 
order to probe the NH region, the two neutrino double-beta decay background events start contributing 
in the total background. Therefore, the detector resolution requires improvement to diminish the contribution 
of 2$\nu\beta\beta$ background events. \par 

The ambiguity of nuclear matrix elements leads to severe uncertainty in the required experimental 
sensitivity. It is observed that using PHFB model the required sensitivity in terms of energy resolution, 
exposure and background rate is in the optimistic scenario in comparison to the SM model. The accurate 
knowledge of the nuclear matrix element is required to minimize the uncertainty in the required 
sensitivity and furthermore, it is the essential parameter for determining the effective mass of 
Majorana neutrino once this 0$\nu\beta\beta$ process is observed.\par

The optimistic region of required sensitivity in terms of the background rate to enter the hierarchy starts from 
$\Lambda$~$\leq$~0.1/tyk and the pessimistic region starts from $\Lambda$~$>$~0.1/tyk for both nuclear 
matrix elements at 3$\sigma$ S.L. and 90\% C.L. Although entering the IH$_{PHFB}$ can tolerate the background 
rate up to $\Lambda$ = 10/tyk at 90\% C.L., for NH$_{PHFB}$ requires $\Lambda$~$\ll$~0.1/tyk. \par

The TIN.TIN experiment at energy resolution 0.5\% at Q$_{\beta\beta}$, needs a minimum exposure of $\Sigma_{min}$~=~0.38 ty to cover the IH$_{PHFB}$ 
completely and in a conservative scenario to cover the IH$_{SM}$ requires $\Sigma_{min}$~=~2.37 ty. Similarly, 
the coverage of NH$_{PHFB}$ requires $\Sigma_{min}$ = 64.05 ty whereas for NH$_{SM}$ this needs~$\Sigma_{min}$~=~3.97$\times10^{2}$~ty.
This $\Sigma_{min}$ is necessity to observe the minimum 1 signal event at background free level. Though 
this $\Sigma_{min}$ is an ideal case, this will provide the limiting factor of the required exposure.

\section*{Acknowledgments}
The authors are grateful to collaborators of the TEXONO Program. 
This work is supported by the Academia Sinica Investigator Award AS-IA-106-M02.
Author~M.~K.~Singh acknowledges University Grant Commission (UGC), India for providing financial support.

\end{document}